\documentstyle[amssymb,aps,preprint]{revtex}

\begin{document}
\draft
\title{A corresponding states approach to Small Angle Scattering from polydisperse
ionic colloidal fluids. }
\author{Domenico Gazzillo, Achille Giacometti}
\address{INFM Unit\`{a} di Venezia and Facolt\`{a} di Scienze, Universit\`{a} di
Venezia, \\
S. Marta DD 2137, I-30123 Venezia, Italy}
\author{Flavio Carsughi}
\address{INFM Unit\`{a} di Ancona and Facolt\`{a} di Agraria, Universit\`{a} di
Ancona,\\
Via Brecce Bianche, I-60131 Ancona, Italy}
\date{\today}
\maketitle


\begin{abstract}
Approximate scattering functions for polydisperse ionic colloidal fluids are
obtained by a corresponding states approach. This assumes that all pair
correlation functions $g_{\alpha \beta }(r)$ of a polydisperse fluid are
conformal to those of an appropriate monodisperse binary fluid (reference
system) and can be generated from them by scaling transformations. The
correspondence law extends to ionic fluids a {\it scaling approximation}
(SA) successfully proposed for nonionic colloids in a recent paper. For the
primitive model of charged hard spheres in a continuum solvent, the partial
structure factors of the monodisperse binary reference system are evaluated
by solving the Orstein-Zernike (OZ) integral equations coupled with an
approximate closure. The SA is first tested within the mean spherical
approximation (MSA) closure, which allows analytical solutions. The results
are found in good overall agreement with exact MSA predictions up to
relevant polidispersity. The SA is shown to be an improvement over the
``decoupling approximation'' extended to the ionic case. The simplicity of
the SA scheme allows its application also when the OZ equations can be
solved only numerically. An example is then given by using the hypernetted
chain (HNC) closure. Shortcomings of the SA approach, its possible use in
the analysis of experimental scattering data and other related points are
also briefly addressed.
\end{abstract}
%
\newpage

\section{Introduction}

Colloidal suspensions of charged particles represent a special class of
ionic fluids \cite{Pusey91,Lowen94,Nagele96}. Unlike solutions of simple
electrolytes such as NaCl, charged colloidal suspensions are highly
asymmetric mixtures, containing both macroions and microions with large size
and charge differences. Moreover, macroions often exhibit
``polydispersity'', which means that particles of a same chemical species
are not necessarily identical, because their size, charge or other
properties may be spread over a large spectrum of values (chemical species
whose particles are all identical are then referred as ``monodisperse'').
The presence of only one polydisperse macroion species is sufficient to make
the colloidal suspension a mixture with a very large number $p$ of
components. The peculiar features of this ``colloidal regime'', namely
asymmetry and polydispersity, give rise to a variety of phenomena concerning
microscopic ordering, phase behaviour, diffusion, and so on.

Experimental information on the structure of such fluids can be obtained
from Small Angle Scattering (SAS) techniques, by using light, neutrons or
X-rays. However, when a significant degree of polydispersity is present in
the sample, the interpretation of experimental data for scattering intensity
is hardly a simple task. In fact, polydispersity and large size-charge
differences represent a serious challenge to the available theoretical
tools. Monte Carlo or molecular dynamics simulations for polydisperse
colloidal fluids involve very large numbers of particles. Moreover, large
size asymmetries at high densities may cause ergodicity problems. On the
other hand, integral equations (IE) of the liquid state theory are
analytically solvable only under special conditions, whereas their numerical
solution for mixtures with large numbers $p$ of components, such as the
polydisperse ones, would require large systems of non-linear equations. As a
consequence, apart from very few peculiar cases \cite{DAguanno91}, IE
numerical studies on multicomponent fluids are usually restricted to $p\ll
10.$ Finally, under the highly demanding conditions of colloidal suspensions
non-convergence problems of the algorithms may often arise.

The present paper will focus on the effects of polydispersity in SAS from
ionic colloidal mixtures, in the framework of IE theories based upon the
Ornstein-Zernike (OZ) equations with approximate closures. Our study refers
to the simplest polydisperse case with only two ionic species: monodisperse
microions and macroions with both size and charge polydispersity. We shall
refer to this system as {\it polydisperse binary }(two-species) ionic
mixture.

To overcome the impossibility of investigating polydisperse systems when IEs
have to be solved numerically, one has to reduce the number of components
and replace the study of a polydisperse fluid with that of a nearly
equivalent but much simpler system. One possibility, not adopted in this
paper, is to neglect microions altogether and approximate the fluid as a
system of macroions interacting through a repulsive screened Coulomb
potential, which implicitly takes into account the contribution of the
neglected particles \cite{Nagele96}. A further refinement of this viewpoint 
\cite{Nagele96,DAguanno92} is to build up an equivalent {\it effective
mixture }with $p^{\prime }\ll p\ $ new components, whose molar fractions and
diameters are determined by replacing a continuous distribution of macroion
sizes with an appropriate $p^{\prime }$-component histogram. Usually, $%
p^{\prime }=$ $3$ is already sufficient and therefore the problem is reduced
to get a numerical solution of IEs for a three-component macroion mixture.
This procedure could be easily extended to include monodisperse microions and
its counterpart would involve a four-component mixture (three for the
macroions plus one for the microions). This method can be expected to be
quite accurate, but it would demand a sizeable amount of numerical work.

In this paper we present an even simpler approach, which requires the
solution of only two-component IEs. We shall show that the problem of a
polydisperse binary ionic mixture can be reduced to the study of a {\it %
monodisperse binary} ionic mixture, with microions and all identical
macroions. The solution for such a reference system is the ``starting''
point for several approximations of increasing accuracy. Our main purpose is
to show that, at the end of this hierarchy, accurate approximate scattering
functions for a polydisperse binary mixture can be easily calculated with
moderate numerical work upon using a corresponding states theory. Our method
hinges on a {\it conformality} \cite{Rowley94} argument, which assumes that
all pair correlation functions of the polydisperse fluid have essentially
the same ``shape'' of their monodisperse binary counterparts and can be
generated from them by means of simple scaling transformations. This
correspondence law is the extension to ionic mixtures of a {\it scaling
approximation }(SA) successfully proposed for nonionic colloids in a recent
paper \cite{Gazzillo99}. This is a non-trivial extension, since the good
performance of the SA for the {\it short-range} potentials of nonionic
colloidal fluids examined in Ref. \cite{Gazzillo99} (uncharged hard sphere
and Lennard-Jones interactions) does not automatically ensures the same
success in the presence of {\it long-range} Coulomb attractions and
repulsions.

To properly treat both macroions and microions on the same footing, the
colloidal suspension will be described by the {\it primitive model} (PM) of
electrolyte solutions, which depicts all ions as charged hard spheres
embedded in a dielectric continuum representing the solvent. The new SA will
be tested, for the PM, against results from an analytic treatment of
polydispersity, which is exact within the {\it mean spherical approximation }%
(MSA) closure for the OZ integral equations. In the PM-MSA case, the OZ
equations were solved analytically many years ago \cite
{Blum75,Blum77,Blum80A,Blum80B}, and, more recently, a closed analytical
formula was obtained for the scattering intensity from charged hard sphere
fluids with any arbitrary number of components \cite{Gazzillo97}. An
essential feature of the SA is that, because of its simplicity, this scheme
can be applied equally well to combinations of potential models and closures
for which only a numerical solution of IEs is possible. It is therefore
possible for instance, as we shall explicitly show, to couple the SA with
the {\it hypernetted chain} (HNC) closure, which is more accurate than the
MSA one for ionic fluids. In these cases the SA becomes a valuable new tool
to predict properties of polydisperse colloidal suspensions in a very simple
way.

The paper is organized as follows. In the next Section the basic formalism
of the small angle scattering and integral equation theory is briefly
recalled along with the primitive model for polydisperse ionic fluid in
Section III. In Section IV our corresponding states treatment of scattering
functions is presented in detail, together with two simpler approximations.
The exact MSA analytical expression for the scattering intensity from
charged hard spheres is also reviewed and some of its predictions for
polydisperse fluids will be reported. In Section V numerical results from
the proposed approximations are compared in detail within the MSA. The
performance of the SA with the HNC closure will then be addressed and few
remarks will be included in the conclusive Section VI.

\section{Small Angle Scattering and Integral Equation Theory}

\subsection{Scattering intensity and structure factors}

An ionic colloidal solution is formed by macroions and microions suspended
in a homogeneous solvent. Usually, this suspending fluid is formed by very
small particles (with respect to the macroions) and is then modelled as a
continuum, characterized by a given dielectric constant and an uniform
density of scattering matter.

According to the scattering theory, the intensity of the scattered radiation
(light, neutrons or X-rays) is proportional to the ensemble or time average,
of $\left| \widetilde{n}({\bf q})\right| ^2$ over all possible equilibrium
configurations of the sample particles. Here ${\bf q}$ is the exchanged wave
vector and $\widetilde{n}({\bf q})$ is the three-dimensional Fourier
transform of $n({\bf r}),$ a quantity related to the density of scattering
matter at the position ${\bf r}$ inside the sample. For neutrons $n({\bf r})$
is the scattering length density: $n({\bf r})=\sum_kb_k\delta ({\bf r-r}_k)$%
, where $b_k$ is the scattering length of the $k-$th nucleus located at $%
{\bf r}_k$ and $\delta $ denotes the Dirac delta function. For X-rays $n(%
{\bf r})$ coincides with the electron density, whereas for light it becomes
the refractive index. In the continuum solvent, $n({\bf r})$ has an uniform
value $n_0$.

In addition to the continuum solvent hypothesis, we assume that inside each
ion (macroion or microion, indifferently) the scattering matter has a
well-defined boundary, i.e., there is a {\it scattering core }with a
well-defined {\it scattering volume,} not necessarily coincident with the
particle volume. While the former depends on the particle-radiation
interaction, the latter is determined by the interparticle repulsions and is
well-defined only in the presence of {\it hard} body repulsions. The
definition of a volume for particles with soft repulsions (e.g.
Lennard-Jones particles) requires in fact some arbitrary and non-universal
convention.

In the case of suspended particles with spherically symmetric interactions
(homogeneous and isotropic fluid) and spherical homogeneous scattering
cores, the SAS theory yields the following expression for the total
scattering intensity $I(q)$ of a $p$-component solution in a volume $V$ \cite
{Nagele96}

\begin{equation}
R(q)\equiv I(q)/V=\rho \sum_{\alpha =1}^p\sum_{\beta =1}^p\sqrt{x_\alpha
x_\beta }F_\alpha (q)F_\beta (q)S_{\alpha \beta }(q),  \label{sas1}
\end{equation}

\noindent as a function of the magnitude of the exchanged wave vector, $%
q\equiv \left( 4\pi /\lambda \right) \sin \left( {\theta /2}\right) ,$ with $%
\lambda $ being the wavelength of the incident radiation and $\theta $ the
scattering angle. The Rayleigh ratio $R(q)$ is the total scattering
intensity per unit volume (also called the differential scattering cross
section and often denoted by ${\frac{d\Sigma }{d\Omega }}(q)$). In Eq. (\ref
{sas1}) $\rho $ is the total number density, while $x_\nu $ and $F_\nu (q)$
are the molar fraction and the form factor of species $\nu $ respectively. 
$F_\nu (q)$ is related to the distribution $n({\bf r})$ of scattering matter
inside particles of species $\nu $ and we can express it as 
\begin{equation}
F_\nu (q)=V_\nu ^{scatt}\left( n_\nu -n_0\right) \frac{3j_1\left( q\sigma
_\nu ^{scatt}/2\right) }{q\sigma _\nu ^{scatt}/2},  \label{sas2}
\end{equation}

\noindent $\sigma _\nu ^{scatt}$ being the diameter of the scattering core$,$
$V_\nu ^{scatt}=\frac \pi 6\left( \sigma _\nu ^{scatt}\right) ^3$ its volume$%
,$ $n_\nu $ the uniform scattering density of species $\nu $ and the
difference $n_\nu -n_0$ its ``contrast'', while $j_1(x)=(\sin x-x\cos x)/x^2$
is the first-order spherical Bessel function. Finally, the functions $%
S_{\alpha \beta }(q)$ are the Ashcroft-Langreth partial structure factors 
\cite{Ashcroft67}

\begin{equation}
S_{\alpha \beta }(q)=\delta _{\alpha \beta }+\rho \sqrt{x_\alpha x_\beta }~%
\widetilde{h}_{\alpha \beta }(q),  \label{sas3}
\end{equation}

\noindent where $\delta _{\alpha \beta }$ is the Kronecker delta and $%
\widetilde{h}_{\alpha \beta }(q)$ is the three-dimensional Fourier transform
of the total correlation function $h_{\alpha \beta }(r)\equiv g_{\alpha
\beta }(r)-1$. Here, $g_{\alpha \beta }(r)$ is the radial distribution
function (RDF) between two particles of species $\alpha $ and $\beta $ at a
distance $r.$

In addition to the scattering intensity, it is then convenient to define a
``measurable'' structure factor \cite{Nagele96} as

\begin{equation}
S_M(q)=\sum_{\alpha =1}^p\sum_{\beta =1}^p\sqrt{x_\alpha x_\beta }w_\alpha
(q)w_\beta (q)\ S_{\alpha \beta }(q),  \label{sm}
\end{equation}

\noindent with weights

\begin{equation}
w_\nu (q)=\frac{F_\nu (q)}{\sqrt{\left\langle F^2(q)\right\rangle }}\ ,
\label{w}
\end{equation}

\noindent the brackets meaning $\left\langle F^2(q)\right\rangle \equiv
\sum_\alpha x_\alpha F_\alpha ^2(q).$ The relationship between $R(q)$ and $%
S_M(q)$ is

\begin{equation}
R(q)=\rho \left\langle F^2(q)\right\rangle S_M(q).  \label{sas6}
\end{equation}

From the theoretical point of view, we will obtain the partial structure
factors $S_{\alpha \beta }(q)$ by solving IEs for the $h_{\alpha \beta }(r).$

\subsection{Integral equations}

The OZ integral equations of the liquid state theory for $p$-component
mixtures with spherically symmetric interparticle potentials are \cite
{Hansen86,Lee88}

\begin{equation}
h_{\alpha \beta }\left( r\right) =c_{\alpha \beta }\left( r\right)
+\sum_{\nu =1}^p\rho _\nu \int d{\bf r}^{\prime }\ c_{\alpha \nu }\left(
r^{\prime }\right) h_{\nu \beta }\left( |{\bf r-r}^{\prime }|\right) ,
\label{as1}
\end{equation}

\noindent where the $c_{\alpha \beta }\left( r\right) $ are the direct
correlation functions and $\rho _\nu \equiv x_\nu \rho $ is the number
density of species $\nu $. These equations can be solved only in combination
with a further relationship between $h_{\alpha \beta }\left( r\right) $ and $%
c_{\alpha \beta }\left( r\right) .$ The formally exact expression of this
``closure'' is

\begin{equation}
c_{\alpha \beta }\left( r\right) =\exp \left[ -(k_BT)^{-1}u_{\alpha \beta
}\left( r\right) +\gamma _{\alpha \beta }\left( r\right) +B_{\alpha \beta
}\left( r\right) \right] -1-\gamma _{\alpha \beta }\left( r\right) ,
\label{as2}
\end{equation}

\noindent where $u_{\alpha \beta }\left( r\right) $ is the interparticle
potential, $k_B$ is Boltzmann's constant, $T$ the absolute temperature, $%
\gamma _{\alpha \beta }\left( r\right) \equiv h_{\alpha \beta }\left(
r\right) -c_{\alpha \beta }\left( r\right) $ and the ``bridge'' functions $%
B_{\alpha \beta }\left( r\right) $ are functionals of $h_{\alpha \beta
}\left( r\right) $ and higher order correlation functions. In practice
however, the exact $B_{\alpha \beta }\left( r\right) $ cannot be calculated,
and several approximations proposed for these functions define a
corresponding series of approximate closures \cite{Hansen86,Lee88}.

The possibility of solving analytically the OZ equations depends on both the
potential model $u_{\alpha \beta }\left( r\right) $ and the chosen closure.
Once that the IEs have been analytically or numerically solved, the partial
structure factors $S_{\alpha \beta }\left( q\right) $ can be obtained from
Eq. (\ref{sas3}).

\section{Primitive model for polydisperse ionic fluids}

We are interested in studying polydispersity effects by properly considering
both macroions and microions on an equal footing. The simplest possibility
is the {\it primitive model }(PM), well known in the theory of electrolyte
solutions. It consists of an electroneutral mixture of $p$ different
components, represented by charged hard spheres embedded in a continuum
solvent of dielectric constant $\varepsilon .$ The species $\alpha ,$ with
diameter $\sigma _\alpha ,$ has molar fraction $x_\alpha $ and electric
charge $z_\alpha e$ ($e$ is the proton charge and $z_\alpha $ the valency).
The interparticle potential $u_{\alpha \beta }\left( r\right) $ is defined by

\begin{equation}
(k_BT)^{-1}u_{\alpha \beta }\left( r\right) =\left\{ 
\begin{array}{l}
+\infty \qquad \qquad \mbox{for\qquad }r<\sigma _{\alpha \beta }\equiv \frac %
12\left( \sigma _\alpha +\sigma _\beta \right) \\ 
\\ 
z_\alpha z_\beta L_B/r\qquad \qquad \mbox{for\qquad }r>\sigma _{\alpha \beta
}
\end{array}
\right.  \label{pm1}
\end{equation}

\noindent where $L_B\equiv e^2/(\varepsilon k_BT)$ is the Bjerrum length.
The electroneutrality condition requires that $\left\langle z\right\rangle
\equiv \sum_{\nu =1}^px_\nu z_\nu =0.$

The PM can also be used for polydisperse colloidal suspensions. In a
``discrete representation'' of polydispersity, a polydisperse two-species
fluid is described by a $p$-component mixture ($p\gg 1$), in which the
monodisperse microions (chemical species 1) are the first component, with
diameter $\sigma _1$, charge $z_1$ and molar fraction $x_1$, while the
remaining $p-1$ components correspond to different varieties of the single
macroion species (chemical species 2). It is often convenient to adopt a
``continuous representation'' of polydispersity, with $p\rightarrow \infty $
and a continuous spectrum of values for the macroion ``disperse'' properties
(size, charge, etc.). In such a {\it continuous-mixture }formalism, we
assume that macroions have a continuous distribution of diameters $\sigma $
around an average one, denoted by $\left\langle \sigma \right\rangle _2.$
For simplicity, we make the further reasonable assumption that the charge
polydispersity of macroions is fully correlated to the size polydispersity.
This can be easily accomplished \cite{Nagele96,DAguanno91} by choosing the
charge (or valency) of each macroion to be proportional to its surface area,
i.e.,

\begin{equation}
z_2(\sigma )=z_{\left\langle \sigma \right\rangle _2}\left( \frac \sigma {%
\left\langle \sigma \right\rangle _2}\right) ^2,  \label{pm2}
\end{equation}

\noindent where $z_{\left\langle \sigma \right\rangle _2}$ is the valency of
the macroions having diameter $\left\langle \sigma \right\rangle _2.$ Both
size and charge distributions of the macroions are therefore governed by a
single independent variable, namely the macroion diameter. The
polydispersity of the macroions can then be expressed by a {\it molar
fraction density function}, $p_2\left( \sigma \right) =x_2^{{\rm tot}%
}f_2(\sigma ),$ where $f_2(\sigma )$ is an appropriate distribution
normalized to unit, while $x_2^{{\rm tot}}=1-x_1$ is the ``amplitude'' of $%
p_2\left( \sigma \right) .$

In the passage from a discrete to a continuous representation of
polydispersity, the molar fractions $x_\nu $ are replaced by $dx=p_2\left(
\sigma \right) d\sigma ,$ the fraction of macroions having diameter in the
range $(\sigma ,\sigma +d\sigma ),$ and the sums $\sum_\nu x_\nu ...$ become
integrals $\int d\sigma \ p_2\left( \sigma \right) ...\ $ The average of a
quantity $Y$ over the macroion distribution is therefore written as

\begin{equation}
\left\langle Y\right\rangle _2\equiv \frac 1{x_2^{{\rm tot}}}\sum_{\nu \in 
{\cal E}_2}x_\nu Y_\nu \longrightarrow \frac 1{x_2^{{\rm tot}}}\int d\sigma
\ p_2\left( \sigma \right) Y\left( \sigma \right) =\int d\sigma \ f_2\left(
\sigma \right) Y(\sigma ),  \label{pm3}
\end{equation}

\noindent where ${\cal E}_2$ denotes the set of indices corresponding to the
macroion components. The average of $Y$ over the whole set of suspended
particles is then: $\left\langle Y\right\rangle \equiv \sum_{\nu =1}^px_\nu
Y_\nu =x_1Y_1+x_2^{{\rm tot}}\left\langle Y\right\rangle _2.$

For $f_2(\sigma )\equiv f(\sigma ;\left\langle \sigma \right\rangle _2,s)$,
we use the Schulz or gamma distribution

\begin{equation}
f(\sigma ;\left\langle \sigma \right\rangle ,s)=\frac{b^a}{\Gamma (a)}\
\sigma ^{a-1}e^{-b\sigma }\;\;(a>1),  \label{pm4}
\end{equation}
\noindent where $\Gamma $ is the gamma function \cite{Abramowitz}, while $%
a=1/s^2,$ $b=a/\left\langle \sigma \right\rangle $ are related to the mean
value $\left\langle \sigma \right\rangle $ and the relative standard
deviation $s\equiv \sqrt{\left\langle \sigma ^2\right\rangle -\left\langle
\sigma \right\rangle ^2}/\left\langle \sigma \right\rangle ,$ which measures
the degree of polydispersity ($0<s<1)$. The choice of the Schulz
distribution is a popular one in colloidal theory because of its
mathematical properties. For $s\rightarrow 0,$ it reduces to a Dirac delta
function centered at $\left\langle \sigma \right\rangle $ (monodisperse
limit). For small values of $s$, $f(\sigma )$ is similar to a Gaussian
distribution, while for larger polydispersity it becomes rather skewed \cite
{Vrij81}. Unlike the Gaussian function, the Schulz distribution is defined
for positive values of $\sigma $ only$.$ Moreover, this distribution allows
a straightforward analytical evaluation of simple averages of the kind
displayed in Eq. (\ref{pm3}). In particular, the first three moments of the
distribution $f_2(\sigma )$ are: $\left\langle \sigma \right\rangle _2,\
\left\langle \sigma ^2\right\rangle _2=\left( 1+s^2\right) \ \left\langle
\sigma \right\rangle _2^2$ and $\left\langle \sigma ^3\right\rangle
_2=\left( 1+s^2\right) \left( 1+2s^2\right) \left\langle \sigma
\right\rangle _2^3,$ while use of Eq. (\ref{pm2}) yields $\left\langle
z\right\rangle _2=z_{\left\langle \sigma \right\rangle _2}\left(
1+s^2\right) .$ These analytical results can be conveniently inserted into
the expressions for the electroneutrality and the packing fraction $\eta $,
i.e.,

\begin{equation}
x_1z_1+x_2^{{\rm tot}}\left\langle z\right\rangle _2=0,  \label{pm5}
\end{equation}

\begin{equation}
\eta =\left( \pi /6\right) \rho \left( x_1\sigma _1^3+x_2^{{\rm tot}%
}\left\langle \sigma ^3\right\rangle _2\right) .  \label{pm6}
\end{equation}

\noindent The microion packing fraction is $\eta _1\equiv \left( \pi
/6\right) \rho _1\sigma _1^3,$ while its macroion counterpart is $\eta
_2\equiv \left( \pi /6\right) \rho _2^{{\rm tot}}\left\langle \sigma
^3\right\rangle _2,$ with $\rho _2^{{\rm tot}}=\rho x_2^{{\rm tot}}.$ From
Eq. (\ref{pm5}) and $x_2^{{\rm tot}}=1-x_1,$ one then gets

\begin{equation}
x_1=\left( 1-\frac{z_1}{\left\langle z\right\rangle _2}\right) ^{-1}=\left[
1-\frac{z_1}{z_{\left\langle \sigma \right\rangle _2}\left( 1+s^2\right) }%
\right] ^{-1},  \label{pm7}
\end{equation}

\noindent which shows that $x_1$ is fully determined by $z_1$ and $%
\left\langle z\right\rangle _2$ ( or equivalently $z_1$, $z_{\left\langle
\sigma \right\rangle _2}$ and $s$).

A final remark is in order. In evaluating the averages of more complex
quantities any analytical integration becomes a formidable or impossible
task and numerical integration brings back to discrete expressions. For this
reason, in the following we shall continue to employ the discrete notation
under the implicit convention that $x_\alpha =x_2^{{\rm tot}}f_2{\bf (}%
\sigma _\alpha {\bf {)}}\Delta \sigma \ $ for the macroion molar fraction ($%
\Delta \sigma \ $is the grid size in the numerical integration).

\section{Approximations and exact expressions}

\subsection{Corresponding states and scaling approximation}

Interparticle potentials are said to be {\it conformal} when they have the
same ``shape'', and systems with conformal interactions are called {\it %
conformal substances }\cite{Rowley94}. Analytically, the {\it conformality }%
of a set of potentials means that all their expressions can be generated
from a single functional form by appropriate scaling of distances and
potential parameters (particle sizes, energies, charges, etc.).

The simplest example refers to pure fluids, when the potential $u_\alpha $
of any species $\alpha ,$ in a set of substances, depends on only two
parameters and can be written as $u_\alpha (r)=\varepsilon _{\alpha \ }%
\widehat{u}\left( r/\sigma _\alpha \right) ,$ where $\sigma _\alpha $ and $%
\varepsilon _{\alpha \ }$are a characteristic length and energy
respectively, while $\widehat{u}$ is a dimensionless function of the
dimensionless ratio $r/\sigma _\alpha .$ The form of $u_\alpha (r)$ implies
that all properties of that set of conformal fluids can be written in terms
of dimensionless reduced variables, and it leads to the ``corresponding
states principle'' commonly found in textbooks \cite{Rowley94}$:$ all
conformal pure fluids at the same dimensionless density and temperature have
identical dimensionless pressure. The RDF of a pure fluid of species $\alpha 
$ in a group of conformal substances can be written as

\begin{equation}
g_\alpha (r;\rho ,T;\sigma _\alpha ,\varepsilon _\alpha )=\widehat{g}\left( 
\frac r{\sigma _\alpha };\rho \sigma _\alpha ^3,\frac{k_BT}{\varepsilon
_\alpha }\right) ,  \label{sa1}
\end{equation}

\noindent where $\widehat{g}$ is a universal function for such a group. If
one among these fluids is arbitrarily chosen as reference system and its
properties are labelled with the subscript 0, then its potential is $%
u_0(r)=\varepsilon _{0\ }\widehat{u}\left( r/\sigma _0\right) $ and its RDF
is given by $g_0(r;\rho ,T;\sigma _0,\varepsilon _0)=\widehat{g}\left(
r/\sigma _0;\rho \sigma _0^3,k_BT/\varepsilon _0\right) .$ From Eq. (\ref
{sa1}) one then gets

\begin{equation}
g_\alpha (r;\rho ,T;\sigma _\alpha ,\varepsilon _\alpha )=g_0(\lambda
_\alpha r\ ;\rho /\lambda _\alpha ^3,T/\xi _\alpha ;\sigma _0,\varepsilon
_0),  \label{sa2}
\end{equation}

\noindent where we have introduced dimensionless scaling factors $\lambda
_\alpha \equiv \sigma _0/\sigma _\alpha $ and $\xi _\alpha \equiv
\varepsilon _\alpha /\varepsilon _0.$ This result is tantamount to say that,
if one knows the RDF of a reference fluid characterized by potential
parameters $\sigma _0,\varepsilon _0,$ then it is possible to derive the RDF
of any conformal fluid of species $\alpha ,$ with potential parameters $%
\sigma _\alpha ,\varepsilon _\alpha .$ The value of $g_\alpha $ at $r$ in a
thermodynamic state ($\rho ,T$) is equal to the value of $g_0$ at the {\it %
scaled} distance $\lambda _\alpha r,$ in the {\it corresponding state }$%
(\rho /\lambda _\alpha ^3,T/\xi _\alpha )$ with {\it scaled} density and
temperature. For instance, if $\sigma _\alpha >\sigma _0$ and $\varepsilon
_\alpha >\varepsilon _0,$ then the corresponding state has a greater density
and a lower temperature. Using the definition of the potential of mean
force, $W\equiv -k_BT\ln g$, Eq. (\ref{sa2}) could also be cast in the form

\begin{equation}
W_\alpha (r;\rho ,T;\sigma _\alpha ,\varepsilon _\alpha )=W_0(\lambda
_\alpha r\ ;\rho /\lambda _\alpha ^3,T/\xi _\alpha ;\sigma _0,\varepsilon
_0).  \label{sa3}
\end{equation}

\noindent For pure fluids then conformality of the potentials implies
conformality of the potentials of mean force and hence of the RDFs. The
potential of mean force between two particles is the sum of the direct pair
potential plus an indirect interaction, due to all the remaining fluid
particles and averaged over all their possible equilibrium configurations.
Finally, a similar property holds true for the structure factors as well

\begin{equation}
S_\alpha (q;\rho ,T;\sigma _\alpha ,\varepsilon _\alpha )=S_0(\lambda
_\alpha ^{-1}q\ ;\rho /\lambda _\alpha ^3,T/\xi _\alpha ;\sigma
_0,\varepsilon _0).  \label{sa4}
\end{equation}

\noindent  The scaling correspondence in $q$-space is that the value of $%
S_\alpha $ at $q$ is equal to the value of $S_0$ at $\lambda _\alpha ^{-1}q$
(in a different thermodynamic state).

On the other hand, for mixtures conformality of potentials does not
necessarily ensure conformality of RDFs in the same simple way. Nevertheless
corresponding states arguments have often been exploited in the liquid state
theory, to postulate approximate conformality relations between mixture and
pure RDFs \cite{Smith73,McDonald73}. Only recently, however, this kind of
approach has been applied to polydisperse fluids and a {\it Scaling
Approximation} (SA) has been proposed for nonionic colloidal suspensions 
\cite{Gazzillo99}. In the SA theory is possible to obtain rather accurate
structure factors for a ``polydisperse one-species'' fluid of uncharged
spherical particles, by first evaluating the RDF $g_{0\text{ }}$of an
appropriate ``monodisperse one-species'' (pure) reference fluid and then
generating all the $p(p+1)/2$ different RDFs of the mixture by taking the
values of the single $g_{0\text{ }}$at suitably scaled distances. The
present work is aimed to extend this SA scheme to polydisperse ionic
colloidal suspensions. It employs two-species fluids with both positive and
negative ions, in order to satisfy the electroneutrality condition. As a
reference system for the ``polydisperse binary'' fluid a suitable
``monodisperse binary'' (M2) mixture is required, where species 1 coincides
with the microions and has their density, size and charge, $(\rho _1^{{\rm %
bin}},\sigma _1^{{\rm bin}},z_1^{{\rm bin}})=(\rho _1,\sigma _1,z_1),$ while
the distribution of macroions is replaced by a single ``average'' component
(species 2) with parameters $(\rho _2^{{\rm bin}},\sigma _2^{{\rm bin}},z_2^{%
{\rm bin}})$. The choice of this reference fluid will be discussed later.
Note that the set of parameters $(\rho _1^{{\rm bin}},\rho _2^{{\rm bin}%
},T;\sigma _1^{{\rm bin}},z_1^{{\rm bin}},\sigma _2^{{\rm bin}},z_2^{{\rm bin%
}})$ can be reduced to $(\rho ^{{\rm bin}},T;\sigma _1,z_1,\sigma _2^{{\rm %
bin}},z_2^{{\rm bin}}),$ since $x_{1\text{ }}^{{\rm bin}}$ is automatically
fixed by the charge ratio through the electroneutrality condition as: $x_{1%
\text{ }}^{{\rm bin}}=\left( 1-z_1/z_2^{{\rm bin}}\right) ^{-1}.$

Our approximation consists in assuming that all RDFs of the polydisperse
ionic mixture are conformal with the RDFs of the monodisperse binary fluid,
which means that

\begin{equation}
~g_{\alpha \beta }\left( r;\rho ,{\bf x}\text{,}T;{\bf \{}\sigma _{\gamma
\delta }\},\{z_{\gamma \delta \ }\}\right) \simeq g_{m_\alpha m_\beta }^{%
{\rm bin}}\left( \lambda _{\alpha \beta }r;\ \rho ,T;\sigma _1,z_1,\sigma
_2^{{\rm bin}},z_2^{{\rm bin}}\right) ,\qquad  \label{sa6}
\end{equation}

\noindent where ${\bf x}$, ${\bf \{}\sigma _{\gamma \delta }\},\{z_{\gamma
\delta \ }\}$ represent the complete set of molar fractions and potential
parameters, $\rho ^{{\rm bin}}=\rho $ , $\lambda _{\alpha \beta }\equiv
\sigma _{m_\alpha m_\beta }^{{\rm bin}}/\sigma _{\alpha \beta },$ with $%
\sigma _{m_{\alpha \ }m_\beta }^{{\rm bin}}\equiv (\sigma _{m_{\alpha \ }}^{%
{\rm bin}}+\sigma _{m_\beta }^{{\rm bin}})/2$, $\alpha ,\beta =1,...,p,~$and

\begin{equation}
\text{ }m_\nu =\left\{ 
\begin{array}{lll}
1 &  & \text{when }\nu =1 \\ 
2 &  & \text{when }\nu \in {\cal E}_2
\end{array}
\right. ,  \label{sa8}
\end{equation}

\noindent  (${\cal E}_2$ was already defined in Eq. (\ref{pm3})). The
correspondence law (\ref{sa6}) provides the recipe for generating all the $%
p(p+1)/2$ independent RDFs of the polydisperse fluid starting from the three
RDFs of the monodisperse binary mixture. It explicitly reads $g_{11}\left(
r\right) \simeq g_{11}^{{\rm bin}}\ (r)$ for microion-microion pairs$,$ $%
g_{1\beta }\left( r\right) \simeq g_{12}^{{\rm bin}}\left( \sigma _{12}^{%
{\rm bin}}\ r/\sigma _{1\beta }\right) ,\ \beta \in {\cal E}_2$ for
microions-macroions, $g_{\alpha \beta }\left( r\right) \simeq g_{22}^{{\rm %
bin}}\left( \sigma _2^{{\rm bin}}\ r/\sigma _{\alpha \beta }\right) ,\
\alpha ,\beta \in {\cal E}_2$ for macroions-macroions.

Our choice of $\lambda _{\alpha \beta }$ for scaling the distances implies
that, when $r<\sigma _{\alpha \beta },$ one gets $r_{\alpha \beta
}^{^{\prime }}\equiv \lambda _{\alpha \beta }r<\sigma _{m_\alpha m_\beta }^{%
{\rm bin}}$ and consequently ensures the correct hard core conditions, $%
g_{\alpha \beta }\left( r\right) =0$ for $r<\sigma _{\alpha \beta }.$ The 
{\it excluded volume }effects, very important for the structure of condensed
phases, are thus properly taken into account by the SA.

Since the Fourier transform of $h_{m_\alpha m_\beta }^{{\rm bin}}(\lambda
_{\alpha \beta }r)$ is $\lambda _{\alpha \beta }^{-3}\ \widetilde{h}%
_{m_\alpha m_\beta }^{{\rm bin}}(\lambda _{\alpha \beta }^{-1}q)$, it is
clear that $S_{m_\alpha m_\beta }^{{\rm bin}}=\delta _{m_\alpha m_\beta
}+\rho ^{{\rm bin}}\sqrt{x_{m_\alpha }^{{\rm bin}}x_{m_\beta }^{{\rm bin}}}~%
\widetilde{h}_{m_\alpha m_\beta }^{{\rm bin}}.$ Under the assumption that $%
\rho ^{{\rm bin}}=\rho $ and upon using Eqs. (\ref{sas3}) and (\ref{sa6})
one then obtains

\begin{equation}
S_{\alpha \beta }(q)^{{\rm SA}}=\delta _{\alpha \beta }+\sqrt{\frac{x_\alpha
x_\beta }{x_{m_\alpha }^{{\rm bin}}x_{m_\beta }^{{\rm bin}}}}~\lambda
_{\alpha \beta }^{-3}\left[ \ S_{m_\alpha m_\beta }^{{\rm bin}}(\lambda
_{\alpha \beta }^{-1}q)-\delta _{m_\alpha m_\beta }\right] \ ,  \label{sas9}
\end{equation}

\noindent where $S_{m_\alpha m_\beta }^{{\rm bin}}\left( q\right) $ is a
shorthand notation for $S_{m_\alpha m_\beta }^{{\rm bin}}\left( q;\ \rho
,T;\sigma _1,z_1,\sigma _2^{{\rm bin}},z_2^{{\rm bin}}\right) ,$ which will
be exploited hereafter unless otherwise specified. Eq. (\ref{sm}), within
this approximation, takes the form

\begin{eqnarray}
S_M(q)^{{\rm SA}} &=&1+x_1w_1^2(q)\left[ S_{11}^{{\rm bin}}(q)-1\right] 
\nonumber \\
&&  \nonumber \\
&&+x_2^{{\rm tot}}\sum_{\alpha \in {\cal E}_2}\sum_{\beta \in {\cal E}_2}%
\frac{x_\alpha }{x_2^{{\rm tot}}}\frac{x_\beta }{x_2^{{\rm tot}}}w_\alpha
(q)w_\beta (q)\left( \frac{\sigma _{\alpha \beta }}{\sigma _2^{{\rm bin}}}%
\right) ^3\left[ \ S_{22}^{{\rm bin}}\left( \frac{\sigma _{\alpha \beta }}{%
\sigma _2^{{\rm bin}}}q\right) -1\right]  \label{sas10} \\
&&  \nonumber \\
&&+2\sqrt{x_1x_2^{{\rm tot}}}~w_1(q)\sum_{\beta \in {\cal E}_2}\frac{x_\beta 
}{x_2^{{\rm tot}}}w_\beta (q)\left( \frac{\sigma _{1\beta }}{\sigma _{12}^{%
{\rm bin}}}\right) ^3\ S_{12}^{{\rm bin}}\left( \frac{\sigma _{1\beta }}{%
\sigma _{12}^{{\rm bin}}}q\right) .  \nonumber
\end{eqnarray}

\noindent 
Eq.(\ref{sas10}) is the basic result of the paper. It provides an expression
for the measurable structure factor of the original polydisperse binary
mixture, once that the partial structure factors of the reference
monodisperse binary mixture are known. In the limit of vanishing charges and
no microions it reduces to the one found in Ref.\cite{Gazzillo99}. The
scattering intensity per unit volume $R(q)^{{\rm SA}}$ is then obtained by
multiplying $S_M(q)^{{\rm SA}}$ by $\rho \left\langle F^2(q)\right\rangle .$

\subsection{Choice of the monodisperse binary mixture}

As reference system, we select a monodisperse 2-component (M2) mixture which
mimics the polydisperse $p$-component fluid. We assume that species 1
coincides with the microions and hence $(\rho _1^{{\rm bin}},\sigma _1^{{\rm %
bin}},z_1^{{\rm bin}})=(\rho _1,\sigma _1,z_1),$ which implies the equality
of the microion packing fraction, i.e., $\eta _1^{{\rm bin}}=\eta _1.$ Then
we replace the polydisperse macroion species, containing $p-1$ components,
with a monodisperse macroion species 2, containing a single ``averaged''
component. To determine its parameters $(\rho _2^{{\rm bin}},\sigma _2^{{\rm %
bin}},z_2^{{\rm bin}})$, we require that

\begin{equation}
\left\{ 
\begin{array}{l}
\rho _2^{{\rm bin}}=\rho _2^{{\rm tot}}, \\ 
\rho _2^{{\rm bin}}\left( \sigma _2^{{\rm bin}}\right) ^3=\rho _2^{{\rm tot}%
}\left\langle \sigma ^3\right\rangle _2\ , \\ 
\rho _1z_1+\rho _2^{{\rm bin}}z_2^{{\rm bin}}=0.
\end{array}
\right.  \label{mb1}
\end{equation}

\noindent The first two equations guarantee that the total number of
macroions and their packing fraction in the M2 mixture are the same as in
the polydisperse fluid ($\eta _2^{{\rm bin}}=\eta _2);$ the third one is the
electroneutrality condition for M2. Combining Eq. (\ref{mb1}) with $\rho _1^{%
{\rm bin}}=\rho _1,$ one finds the solution 
\begin{equation}
\left\{ 
\begin{array}{l}
\rho ^{{\rm bin}}=\rho ,\qquad \text{and}\qquad x_1^{{\rm bin}}=x_1, \\ 
\sigma _2^{{\rm bin}}=\left\langle \sigma \right\rangle _2^{1/3}, \\ 
z_2^{{\rm bin}}=\left\langle z\right\rangle _2\ .
\end{array}
\right.   \label{mb2}
\end{equation}

\noindent In this way, the definition of the set of M2 parameters $(\rho ^{%
{\rm bin}},T;\sigma _1,z_1,\sigma _2^{{\rm bin}},z_2^{{\rm bin}})$ is
complete.

Choices other than (\ref{mb2}) are clearly possible. We have explicitly
worked out few of them and found that they do not significantly alter the
final numerical results. Eq. (\ref{mb2}) has then been privileged on the
basis of its simplicity and natural physical interpretation.

Besides being used as a reference system for SA, the M2 mixture may itself
be regarded as the simplest approximation to the polydisperse $p$-component
fluid. The corresponding measurable structure factor would then be

\begin{eqnarray}
S_M(q)^{{\rm M2}} &=&1+x_1w_1^2(q)\left[ S_{11}^{{\rm bin}}(q)-1\right]
+~x_2^{{\rm tot}}[w_2^{{\rm bin}}(q)]^2~\left[ \ S_{22}^{{\rm bin}}\left(
q\right) -1\right]  \nonumber \\
&&  \label{mb4} \\
&&+2\sqrt{x_1x_2^{{\rm tot}}}~w_1(q)\ w_2^{{\rm bin}}(q)S_{12}^{{\rm bin}%
}\left( q\right) .  \nonumber
\end{eqnarray}
\noindent which simply corresponds to approximate the original polydisperse
binary mixture with a plain monodisperse binary mixture.

\subsection{Extended decoupling approximation}

To emphasize the role played by the scaling of distances in the SA, let us
consider the simpler case of no scaling. This can be obtained from the SA
expressions by setting $\lambda _{\alpha \beta }=1$ everywhere. The result
corresponds to an approximation which provides an exact evaluation of all
form factors of the polydisperse system but assumes that the RDFs can be
replaced by a set of only three effective RDFs of a monodisperse binary
ionic fluid. Hence, in the previous language we have $g_{11}\left( r\right)
\simeq g_{11}^{{\rm bin}}\left( \ r\right) ,$ $g_{1\beta }\left( r\right)
\simeq g_{12}^{{\rm bin}}\left( r\right) $ and $g_{\alpha \beta }\left(
r\right) \simeq g_{22}^{{\rm bin}}\left( r\right) \ $for micro-micro,
micro-macro and macro-macro ionic pairs, respectively$.$ Eq. (\ref{sas10})
simplifies to

\begin{eqnarray}
S_M(q)^{{\rm EDA}} &=&1+x_1w_1^2(q)\left[ S_{11}^{{\rm bin}}(q)-1\right]
+~x_2^{{\rm tot}}\left\langle w(q)\right\rangle _2^2~\left[ \ S_{22}^{{\rm %
bin}}\left( q\right) -1\right]  \nonumber \\
&& \\
&&+2\sqrt{x_1x_2^{{\rm tot}}}~w_1(q)\ \left\langle w(q)\right\rangle
_2~S_{12}^{{\rm bin}}\left( q\right) .  \nonumber
\end{eqnarray}

\noindent The superscript EDA means {\it Extended Decoupling Approximation, }%
since this approximation may be reckoned as an extension to polydisperse
ionic colloids of the ``decoupling approximation'' (DA), proposed by
Kotlarchyk and Chen \cite{Kotlarchyk83} for nonionic fluids and well known
to the small angle scattering experimentalists. The EDA may also be regarded
as a special limiting case of the {\it binary substitutional model} proposed
by N\"{a}gele {\it et al.} \cite{Nagele93} for a different colloidal model
with two polydisperse macroion species and no microions.

\subsection{MSA closure and analytic expressions}

The expressions we have previously derived for SA, EDA and M2 are clearly
independent of the approximate ``closure'' chosen for solving the OZ
equations. One then expects that an improvement in the selection of the
closure would provide increasingly accurate results for the polydisperse
colloidal suspension. In the present paper we shall focus mainly on the MSA
closure, to take advantage of its analytical properties. Another example
will be considered in Section V-C.

For the PM, the MSA consists in adding to the exact hard sphere condition, $%
g_{\alpha \beta }\left( r\right) =0$ or $h_{\alpha \beta }\left( r\right)
=-1 $ when $r<\sigma _{\alpha \beta },$ the approximate relationship
(closure) 
\begin{equation}
c_{\alpha \beta }\left( r\right) =-(k_BT)^{-1}u_{\alpha \beta }(r)\qquad
\qquad \mbox{for\qquad }r>\sigma _{\alpha \beta },  \label{msa1}
\end{equation}

\noindent which is asymptotically correct for $r\rightarrow \infty .$ The
advantage of the MSA closure is that the corresponding OZ equations for the
PM were solved analytically some times ago \cite{Blum75}. Senatore and Blum 
\cite{Senatore85} employed MSA expressions for the partial structure factors 
$S_{\alpha \beta }(q)$ to calculate numerically $S_M(q)$ for charged hard
spheres with {\it either} size polydispersity {\it or} charge
polydispersity. More recently, a {\it closed} MSA formula for $S_M(q)$
by-passing the explicit calculation of the partial structure factors was
obtained in Ref. \cite{Gazzillo97}. This was the extension to ionic systems
of an analogous expression for polydisperse uncharged hard spheres in the PY
approximation \cite{Vrij79}. For the sake of completeness, the MSA analytic
expression for the scattering intensity is reported in Appendix where some
misprints appearing in Ref.\cite{Gazzillo97} are also corrected. The MSA
closure yields analytic expressions for both $S_{m_\alpha m_\beta }^{{\rm bin%
}}\left( q\right) $ and $S_M(q)$ depending on a single screening parameter $%
2\Gamma ,$ which in turn has to be determined self-consistently.

A well known drawback of the MSA is that, for dilute solutions of highly
charged particles, it may predict unphysical negative values for $g_{\alpha
\beta }\left( r\right) $ near the contact distance $\sigma _{\alpha \beta }$
or in a neighborhood of the first minimum. Some proposal have been advanced
to heal this restriction \cite{Hansen82,Belloni86,Ruiz90}. For simplicity
however, the emphasis of the present work will be mainly on concentrated
suspensions of weakly charged particles. In this regime the MSA is
reasonably accurate, with the Coulomb part of the potential being only a
perturbation with respect to the hard sphere one. The above remark is
nevertheless by no means a limit to our method which could be easily
associated to more accurate closures such as the ``hypernetted chain
approximation'' (HNC), corresponding to take $B_{\alpha \beta }\left(
r\right) =0$ in Eq. (\ref{as2}), or the self-consistent mixing scheme (HMSA)
proposed by Zerah and Hansen for potentials with attractive terms \cite
{Hansen86,Lee88}. Clearly, in the HNC or HMSA integral equations the
monodisperse binary reference fluid can be treated only numerically. This
point will be further discussed in Section V-C where an example of such
calculation will be provided for the HNC.

\section{Numerical calculations}

In order to display the behavior of the scattering functions under some
typical polydisperse conditions, we numerically reproduced a realistic
experimental environment.

The microions were given a valency $z_1=$ $+1$ and a diameter $\sigma _1=$ $5
$ \AA\ (solvated counterions), while we used an average macroion size $%
\left\langle \sigma \right\rangle _2=100$ \AA\ with relatively small charges 
$z_{\left\langle \sigma \right\rangle _2}$ in the range $-20\div 0$ (in $e$
units) to ensure meaningful MSA results as previously discussed. We will
increase this value up to $z_{\left\langle \sigma \right\rangle _2}=-50$
later on using HNC. The scattering due to the microions is in principle not
completely negligible and it might be also characterized by a different
contrast with respect to the macroions. Nevertheless both contrasts were
here fixed to the same value $\Delta n=4\times 10^{10}$ ${\rm cm}^{-2}$,
which is typically found in neutron scattering from silica particles
suspended in water \cite{Kline96}. In evaluating the form factors we further
assumed $\sigma _\nu ^{scatt}=\sigma _\nu $ for all particles. A room
temperature $T=298$ K and the dielectric constant $\varepsilon =78$ of water
result into a value $L_B=7.189$ \AA\ for the Bjerrum length. All numerical
calculations were performed for packing fractions $\eta =0.1,$ $0.3$ and
polydispersity $s=0,$ $0.1,$ $0.2$ and $0.3$ (the first value corresponding
to the monodisperse binary mixture). We note that when $s=0$, it is
necessary to take $\left| z_2^{{\rm bin}}\right| \lesssim 30$ for $\eta =$ $%
0.3$ and $\left| z_2^{{\rm bin}}\right| \lesssim 10$ for $\eta =$ $0.1,$ to
avoid unphysical negative values of the MSA $g_{22}^{{\rm bin}}$ at contact.

The three Schulz distributions, with polydispersity $s=0.1,$ $0.2$ and $0.3$%
, were discretized with a grid size $\Delta \sigma /\left\langle \sigma
\right\rangle _2=0.02$, and truncated at $\sigma _{{\rm cut}}/\left\langle
\sigma \right\rangle _2=1.56$, $2.22$ and $2.96,$ respectively. These $%
\sigma _{{\rm cut}}$ values correspond to polydisperse mixtures with a
number of macroion components equal to $79$, $112$ and $149$, practically
intractable with the available algorithms for solving IEs numerically.

\subsection{Polydispersity and charge effects in exact MSA results}

Before analyzing the performance of the SA and other approximations, it is
useful to recall how size and charges polydispersity affect the measurable
scattering structure factor. This is achieved by using the closed analytical
expression for $S_M(q)$ which is given in Appendix which is exact within the
MSA.

Fig. 1 depicts the effects of polydispersity on the measurable structure
factor. $S_M{}(q)^{{\rm MSA}}$ is plotted as a function of the dimensionless
variable $q\left\langle \sigma \right\rangle _2$ for increasing values of $s$
and fixed $\eta =0.3$ and $z_{\left\langle \sigma \right\rangle _2}=-20.$ We
note that as $s$ increases at fixed $\eta$, $\rho$ decreases. As expected,
the effect of increasing polydispersity is three-fold: i) the oscillations
on the tail of the curves are greatly reduced as a consequence of the
destructive interference stemming from the several length scales involved;
ii) the first peak is lowered, broadened and shifted to smaller $q$ values
corresponding to a larger typical distance between macroions-macroions
nearest-pairs; iii) the $q\rightarrow 0$ limit is increased since highly
dispersed particles can be more efficiently packed. All these effects
parallel those observed in polydisperse nonionic fluids \cite
{Gazzillo99,Vrij81} as well as in mixtures constituted of only macroions
interacting through a repulsive screened Coulomb interactions \cite
{DAguanno90}, and they were already recorded even in the PM \cite{Senatore85}%
.

Next we check the effect of the charge. This is reported in Fig. 2, where
the $S_M{}(q)^{{\rm MSA}}$ corresponding to $\eta =0.3$ , $z_{\left\langle
\sigma \right\rangle _2}=-20$, $s=0.3$ is compared with that of the
polydisperse mixture of neutral hard spheres which results from ``switching
off'' all charges and leaving all other parameters unchanged$.$ As the
charge increases, the main peak becomes higher and shifts to smaller $q$
values, since its position is essentially determined by the macroion-macroion
equilibrium distance which becomes larger in the presence of electrostatic
repulsions. The difference in the $q\rightarrow 0$ behavior is also
evident: the charges lower the $S_M(q)$ values near the origin, as a
consequence of the long-range nature of the Coulomb potential.

\subsection{Scaling approximation plus MSA}

Our aim is now to display the performance of the SA when the partial
structure factors $S_{m_\alpha m_\beta }^{{\rm bin}}(q)$ of the reference M2
mixture are evaluated using the MSA closure. These SA-MSA results are
compared with the exact MSA solution for polydisperse charged hard spheres
previously discussed. We shall comment in the next subsection on a method we
envisaged to avoid unnecessary repetitions in the IE calculations for the M2
mixture.

In Fig. 3 the structure factor $S_M(q)^{{\rm SA-MSA}}$ is shown for two
different degrees of polydispersity ($s=0.1$ and $0.3$) at low concentration
and under weak charge conditions ($\eta =0.1$ and $z_{\left\langle \sigma
\right\rangle _2}=-10$). The corresponding results from the M2-MSA and
EDA-MSA approximations are also reported for comparison. As expected, at
small polydispersity (Fig. 3a) there are very little differences among all
these curves, although the EDA yields a somewhat larger value for $S_M(q=0).$
This overestimation of the low-$q$ scattering in the EDA becomes much larger
as polydispersity increases (Fig. 3b). This is the same qualitative trend
resulted in the DA for neutral systems 
\cite{Pusey91,Nagele96,Gazzillo99,Pedersen94}.
At $s=0.3$ it is apparent that the position of the first peak in the EDA
follows that of the M2, whose maximum is shifted to larger $q$ values with
respect to the MSA result. On the other hand, the SA reproduces more
accurately the position of the first peak and follows very closely the
correct curve for $q\left\langle \sigma \right\rangle _2\gtrsim 6.$

The discrepancy in the low-$q$ region, which is in fact the most interesting
from the SAS point of view, can be more clearly seen in Fig. 4, where the
scattering intensity per unit volume $R(q)$, calculated for the same
parameters of Fig. 3 is displayed on a log-log scale. Nevertheless the SA
performs overall rather well in all regions.

In Fig. 5 the same functions of Fig. 3 are then plotted for a more
concentrated suspension and higher macroion charges ($\eta =0.3$ and $%
z_{\left\langle \sigma \right\rangle _2}=-20$), again for $s=0.1$ and $0.3$,
while the corresponding results for $R(q)$ are displayed in Fig. 6. Figs. 5a
and 5b yield compelling evidence of the potentiality of the SA confirming
the previous remarks. It is useful to analyse these results in the sequence $%
{\rm M2\rightarrow EDA\rightarrow SA.}$ In Fig. 5a $S_M(q)^{{\rm EDA}}$
exactly coincides with $S_M(q)^{{\rm M2}}$ in the first peak region, but
differs from it at the locations of the M2 minima and at small $q$ values.
On the contrary, the SA curve is close to the exact MSA one everywhere. In
Fig. 5b the M2 approximation largely disagrees with the MSA one. The EDA
improves here on the M2 since it takes all form factors of the
polydisperse fluid correctly into account. It exhibits a lower peak height
and practically no subsequent oscillations. Nevertheless, the EDA has a
dramatic low-$q$ overestimate and it behaves poorly in essentially all regions.
On the contrary, the SA is fairly accurate in the whole experimentally
accessible $q$-range. Its performance at $\eta =0.3$ appears to be even more
accurate then at $\eta =0.1.$ All these features are quite remarkable if we
recall that the SA, EDA and M2 curves have been obtained starting from the
same partial structure factors $S_{m_\alpha m_\beta }^{{\rm bin}}(q).$ This
fact clearly shows the crucial role played by the scaling of the distances.
Its effect is to shift the first peak position to the right location and to
dump all oscillations after the first peak. Physically it confirm the
soundness of our ``conformality'' hypothesis as expressed by Eq. (\ref{sa6})
and it shows that overlooking differences among macroion-macroion RDFs at
contact (as it is done in the SA) is a reasonable assumption.

\subsection{Scaling approximation plus HNC}

To illustrate the possibility of applying the SA scheme even when the OZ
equations admit only numerical solutions, we investigated SA with the HNC
closure and analyzed two cases: $\eta =0.1$, $s=0.3$ with $z_{\left\langle
\sigma \right\rangle _2}=-10$ and $z_{\left\langle \sigma \right\rangle
_2}=-50$, respectively (all other parameters were fixed as before). While
the first case was already studied with the SA-MSA, the second one
represents a situation, of low concentration and high charges, in which the
MSA yields negative values of the macroion-macroion RDF at contact and hence
cannot be utilized. This drawback is avoided by using the HNC closure.

In both cases we solved the HNC equations for the parameter values of the
corresponding M2 reference mixture, using an $r$-space grid size $\Delta
r/\left\langle \sigma \right\rangle _2=0.02$ and a number of grid points $%
{\cal N}=4096.$ This choice implies that $q_{_{{\rm \max }}}\left\langle
\sigma \right\rangle _2=50\pi ,$ with a small enough grid size in $q$-space, 
$\Delta q=q_{{\rm \max }}/{\cal N}$, allowing the implementation of a
``trick'' proposed in Ref. \cite{Gazzillo99} . In fact, the expression for $%
S_M(q)^{{\rm SA}}$, Eq. (\ref{sas10}), would require, at each $q,$ the
evaluation of one term $S_{11}^{{\rm bin}}(q)$, $p-1$ terms $\ S_{12}^{{\rm %
bin}}\left( q\sigma _{1\beta }/\sigma _{12}^{{\rm bin}}\right) $ and $%
(p-1)p/2$ terms $S_{22}^{{\rm bin}}\left( q\sigma _{\alpha \beta }/\sigma
_2^{{\rm bin}}\right) $ (recall that $p-1$ is the number of macroion
components). These cumbersome repeated calculations can be avoided. We
calculated $S_{11}^{{\rm bin}}$,$\ S_{12}^{{\rm bin}}$ and $S_{22}^{{\rm bin}%
}$ at the grid points $q_i=i\Delta q$ ($i=0,\ldots ,{\cal N}-1$) {\it only
once}, storing all these values in arrays. Although the grid points $q_i$ do
not exhaust the whole set of $q\sigma _{1\beta }/\sigma _{12}^{{\rm bin}}$
and $q\sigma _{\alpha \beta }/\sigma _2^{{\rm bin}}$ values required in Eq. (%
\ref{sas10}), the stored structure factors represent a fine sampling of
these continuos functions. Therefore, if the $\Delta q$ is small enough, the
value of $S_{12}^{{\rm bin}}$ (or $S_{22}^{{\rm bin}})$ at a certain point
can be approximated with that at the nearest grid point with a negligible
error. In this way the sums of Eq. (\ref{sas10}) can be quickly performed.

Figure 7 shows the $S_M(q)^{{\rm SA-HNC}}$ curves, along with the SA-MSA one
for $z_{\left\langle \sigma \right\rangle _2}=-10.$ As expected, in the
lower charge case the SA-HNC prediction is very close to the SA-MSA one,
with only a slight shift in the first peak position. The SA-HNC structure
factor with $z_{\left\langle \sigma \right\rangle _2}=-50$ is a not trivial
result: it refers to a polydisperse colloidal system with mean size ratio $%
\sigma _1:\left\langle \sigma \right\rangle _2=$1:20 and mean (absolute)
charge ratio $\left| z_1\right| :\left| z_{\left\langle \sigma \right\rangle
_2}\right| =$1:50. Unfortunately, in this case we cannot make a comparison
with ``exact'' data. On the other hand, the lack of these data and the
difficulty of generating them in a very asymmetric regime is just the
strongest motivation for introducing approximate theories such as the SA.

\section{Concluding remarks}

In this paper the problem of computing scattering functions for polydisperse
ionic colloidal fluids has been addressed by integral equation methods. In
the framework of the primitive model we have shown that, despite the
complexity of these systems, surprisingly accurate predictions can be
obtained with a limited numerical effort. We have successfully extended the 
{\it scaling approximation }introduced in Ref. \cite{Gazzillo99} for
polydisperse fluids of neutral particles. The SA still works well when
Coulombic (both repulsive and attractive) interactions are present,
notwithstanding the strong charge-size asymmetries of the polydisperse
colloidal regime. Only the study of an appropriate monodisperse binary
mixture (the M2 reference system) is required for a complete
characterization of the polydisperse system.

Our corresponding states-like theory is based on the simple physical idea of
conformality of all RDFs in the polydisperse mixture. All partial structure
factors are generated by scaling their three counterparts of the M2 fluid.
In the liquid state theory similar ideas have been widely exploited in the
past \cite{Smith73,McDonald73} but, to our knowledge, Ref. \cite{Gazzillo99}
and this paper are the first application to polydisperse fluids.

Clearly, the SA theory is accurate only in the average. In fact, the scaling
is hardly accurate for each individual pair correlation $g_{\alpha \beta }(r)
$. In particular, it incorrectly assumes the equality of the RDFs at contact
for all macroion-macroion pairs as well as for the microion-macroion ones.
However, an essential feature of our SA is that it correctly ensures $%
g_{\alpha \beta }(r)=0$ inside the hard cores. These excluded volume
conditions are crucial, as it is shown by the failure of the ``extended
decoupling approximation'' which neglects them. The structure factors $%
S_M(q)^{{\rm SA}}$ turn out to be accurate in the first peak region and
beyond; some inaccuracy, due to the harsh approximations of our theory, is
found at low $q$ values. Since the $q\rightarrow 0$ limit is related to
thermodynamics, this means that the SA can be meaningfully exploited to
extract structural but not thermodynamical predictions.

Because of its simplicity, the SA can be safely employed also when the OZ
integral equations have to be solved numerically since its application to
both different closures and different potential for ionic colloids is
feasible. These features are indicative that SA is a useful theoretical tool
to investigate, to first approximation, the structure of polydisperse
(nonionic and ionic) colloids under highly demanding conditions. The
existence of a good approximation which reduces the study of polydisperse
fluids to that of an effective monodisperse one should not be
underestimated. Real life colloids are always polydisperse to a certain
degree and polydispersity always represents a challenge in the
interpretation of experimental data. We hope that the SA will result
particularly useful in the analysis of Small Angle Scattering data, since it
considerably outperforms the ``decoupling approximation'', popular in this
context, at the cost of a minimal additional effort.

It would be interesting to compare our theory with the approach proposed by
D'Aguanno and Klein \cite{DAguanno92}. As already mentioned in the
Introduction, these authors followed a different point of view and replaced
the continuous Schulz distribution with an histogram containing a finite
number of well-chosen diameters, thus reducing the polydisperse fluid to an
effective mixture with a very small number $p^{\prime }$ of components. In
this case however we expect a non-trivial increase in the numerical effort
involved. In fact, to avoid the rapid increase in computational cost with
increasing $p^{\prime }$ in the D'Aguanno-Klein mixtures, that approach was
recently modified by Lado and coworkers by adding an orthogonal polynomial
expansion technique \cite{Lado96}, and afterwards by merging this with a
thermodynamic perturbation scheme \cite{Leroch99}. A comparison of the SA
with these alternative theories is left to future work.

\acknowledgments
Financial support from the Italian MURST (Ministero dell'Universit\`{a} e
della Ricerca Scientifica e Tecnologica) through the INFM (Istituto
Nazionale di Fisica della Materia) is gratefully acknowledged. One of us
(DG) thanks Giorgio Pastore (Trieste) and Enrique Lomba (Madrid) for
providing their HNC computer codes.

\appendix

\section{MSA expression for the scattering intensity}

In this appendix we report the basic formulas involved in the MSA
calculation of the scattering intensity from charged hard sphere fluids, as
described in \cite{Gazzillo97}. Let us introduce the following short-hand
notations:

\begin{equation}
\left\{ Y\right\} _0\equiv \sum_{\nu =1}^p\rho _\nu Y_\nu =\rho \left\langle
Y\right\rangle ,
\end{equation}
\begin{equation}
\left\{ Y\right\} \equiv \sum_{\nu =1}^p\rho _\nu Y_\nu e^{iX_\nu }=\rho
\left\langle Ye^{iX}\right\rangle ,
\end{equation}
where $X_\nu \equiv q\sigma _\nu /2{.}$ The MSA analytical solution depends
on the screening parameter $2\Gamma ,$ which must be determined numerically
by solving the consistency equation

\begin{equation}
\left( 2\Gamma \right) ^2=4\pi L_B\sum_{\nu =1}^p\rho _\nu \left( \frac{%
z_\nu -P_z\sigma _\nu ^2/2}{1+\Gamma \sigma _\nu }\right) ^2,
\end{equation}

\noindent where

\begin{equation}
P_z=\frac \pi \Omega \left\{ {\frac{\sigma z}{1+\Gamma \sigma }}\right\} _0\
,  \label{pz}
\end{equation}

\begin{equation}
\Omega =\Delta +{\frac \pi 2}\left\{ {\frac{\sigma ^3}{1+\Gamma \sigma }}%
\right\} _0\ ,
\end{equation}

\noindent with $\Delta =1-\eta .$ These quantities are also required to
compute

\begin{equation}
A_\nu =\frac{L_B}\Gamma \frac{z_\nu -P_z\sigma _\nu ^2/2}{1+\Gamma \sigma
_\nu }.
\end{equation}
\noindent In the limit of point ions (all $\sigma _\nu \rightarrow 0$), $%
2\Gamma $ becomes the Debye inverse shielding length $\kappa _D$ of the
Debye-H\"{u}ckel theory for electrolyte solutions, while for finite size
ions it is always a lower bound (i.e. $2\Gamma \leq \kappa _D$). We also need

\begin{equation}
\alpha _\nu ={\frac{\pi \sigma _\nu ^3\phi _\nu }{6\Delta },}
\end{equation}
\begin{equation}
\beta _\nu ={\frac{\pi \sigma _\nu ^2\psi _\nu }{2\Delta },}
\end{equation}
\begin{equation}
\gamma _\nu ^{(0)}=\frac{2\pi i}q\frac{\Gamma ^2}{L_B}\ A_\nu \sigma _\nu
\psi _\nu \ {,}
\end{equation}
\begin{equation}
\gamma _\nu ^{(1)}=\frac{2\pi i}q\frac \Gamma {L_B}\ A_\nu e^{-iX_\nu },
\end{equation}
\begin{equation}
\gamma _\nu =\gamma _\nu ^{(0)}+\gamma _\nu ^{(1)},
\end{equation}

\noindent where $\psi _\nu =j_0(X_\nu )$ and $\phi _\nu =3j_1(X_\nu )/X_\nu
, $ with $j_0(x)=\sin x/x$ and $j_1(x)=(\cos x-x\cos x)/x^2$ being Bessel
functions.

The final expression for the scattering intensity per unit of volume is 
\begin{equation}
R(q)=R_1(q)+R_2(q),
\end{equation}
\noindent where 
\begin{eqnarray}
R_1 &=&\left\{ F^2\right\} _0+\left\{ \alpha ^2\right\} _0\left| c_1\right|
^2+\left\{ \beta ^2\right\} _0\left| c_2\right| ^2 \\
&+&2{\rm Re}\left[ \ \left\{ F\alpha \right\} _0c_1+\left\{ F\beta \right\}
_0c_2+\left\{ \alpha \beta \right\} _0c_1c_2^{*}\ \right] \ ,  \nonumber
\end{eqnarray}
\begin{equation}
R_2=\left\{ \left| q\gamma \right| ^2\right\} _0\left| c_3\right| ^2+2{\rm Re%
}\left[ \ \left\{ q\gamma F\right\} _0\ c_3+\left\{ q\gamma \alpha \right\}
_0\ c_3c_1^{*}+\left\{ q\gamma \beta \right\} _0\ c_3c_2^{*}\ \right] .
\end{equation}

\noindent Here, $F_\nu $ is the form factor given by Eq. (\ref{sas2}), ${\rm %
Re}\left[ \ \ldots \right] $ the real part of a complex number and the
asterisk denotes complex conjugation. Other necessary quantities appearing
in these equations are

\begin{equation}
c_1=\frac{t_2}{t_1},\qquad c_2=\frac{t_3}{t_1},\qquad c_3=\frac{t_4}{t_1},
\end{equation}

\noindent $t_m$ ($m=1,\ldots ,4$) being the cofactor of the ($1,m$)th
element of the first row in the following determinant

\begin{equation}
\left| 
\begin{array}{llll}
~\rho _\nu ^{1/2}F_\nu & \qquad \rho _\nu ^{1/2}\alpha _\nu & \ \qquad \rho
_\nu ^{1/2}\beta _\nu & \quad \quad \rho _\nu ^{1/2}q\gamma _\nu \\ 
\left\{ F\right\} & \quad 1+\ \left\{ \alpha \right\} & \quad \{\beta
\}-3\xi _2/\Delta +iq/2 & \quad \{q\gamma ^{(0)}\}-2i\Gamma P_z\Delta \\ 
\left\{ \sigma F\right\} & \qquad \left\{ \sigma \alpha \right\} & \quad
1+\{\sigma \beta \} & \qquad \{\sigma q\gamma ^{(0)}\} \\ 
\left\{ AF\right\} & \qquad \left\{ A\alpha \right\} & \qquad \left\{ A\beta
\right\} & \quad q+\{Aq\gamma ^{(0)}\}+2i\Gamma
\end{array}
\right| ,  \label{det}
\end{equation}

\noindent and where $\xi _2=(\pi /6)\left\{ \sigma ^2\right\} _0$ .

Eqs.(\ref{pz}) and (\ref{det}) correct the misprints appearing in the
corresponding equations of Ref. \cite{Gazzillo97}. In the expression of $P_z$
given by Eq. (48) of that paper the factor $\pi $ was omitted. We also note
that our definition of $P_z$ and $\Omega $ differs from Blum's original one 
\cite{Blum75}. 
\begin{figure}[tbp]
\caption{Polydispersity effects. Exact MSA structure factor $S_M(q)$ of
polydisperse charged hard spheres, at fixed packing fraction $\eta =0.3$,
for different degrees of polydispersity $s$, ($s=0$ corresponds to the
monodisperse binary case). Other parameters: $\sigma _1= 5 $ \AA , $z_1 = +1$
($e$ units) for microions; $\left\langle \sigma \right\rangle _2 =100$ \AA , 
$z_{\left\langle \sigma \right\rangle _2}= -20$ for macroions.}
\end{figure}

\begin{figure}[tbp]
\caption{Charge effects. The exact MSA structure factor $S_M(q)$ for
polydisperse charged hard spheres with $\eta =0.3$, $z_{\left\langle \sigma
\right\rangle _2}= -20$, $s = 0.3$ (other parameters as in Figure 1) is
compared with its exact PY counterpart for the corresponding polydisperse
mixture of neutral hard spheres.}
\end{figure}

\begin{figure}[tbp]
\caption{(a) Structure factor $S_M(q)$ for $\eta =0.1$, $z_{\left\langle
\sigma \right\rangle _2}= -10$, $s = 0.1$ (other parameters as in Figure 1).
Comparison of M2-MSA, EDA-MSA, SA-MSA and exact MSA results. (b) Same as
(a), but for $s = 0.3$.}
\end{figure}

\begin{figure}[tbp]
\caption{Scattering intensity per unit volume, $R(q)$, using a log-log
scale. Comparison of M2-MSA, EDA-MSA, SA-MSA and exact MSA results. The
systems are the same as in Figure 3: (a) $\eta =0.1$, $z_{\left\langle
\sigma \right\rangle _2}= -10$, $s = 0.1$. (b) Same as (a), but for $s=0.3$.}
\end{figure}

\begin{figure}[tbp]
\caption{(a) Structure factor $S_M(q)$ for $\eta =0.3$, $z_{\left\langle
\sigma \right\rangle _2}= -20$, $s = 0.1$ (other parameters as in Figure 1).
Comparison of M2-MSA, EDA-MSA, SA-MSA and exact MSA results. (b) Same as
(a), but for $s = 0.3$.}
\end{figure}

\begin{figure}[tbp]
\caption{Scattering intensity per unit volume, $R(q)$, using a log-log
scale. Comparison of M2-MSA, EDA-MSA, SA-MSA and exact MSA results. The
systems are the same as in Figure 5: (a) $\eta =0.3$, $z_{\left\langle
\sigma \right\rangle _2}= -20$, $s = 0.1$. (b) Same as (a), but for $s=0.3$.}
\end{figure}

\begin{figure}[tbp]
\caption{SA-HNC predictions for the structure factor $S_M(q)$ at $\eta =0.1$%
, $s=0.3$, with $z_{\left\langle \sigma \right\rangle _2}= -10$ and $%
z_{\left\langle \sigma \right\rangle _2}= -50$ (other parameters as in
Figure 1). In the $z_{\left\langle \sigma \right\rangle _2}= -10$ case the
corresponding SA-MSA curve is also plotted for comparison.}
\end{figure}


%

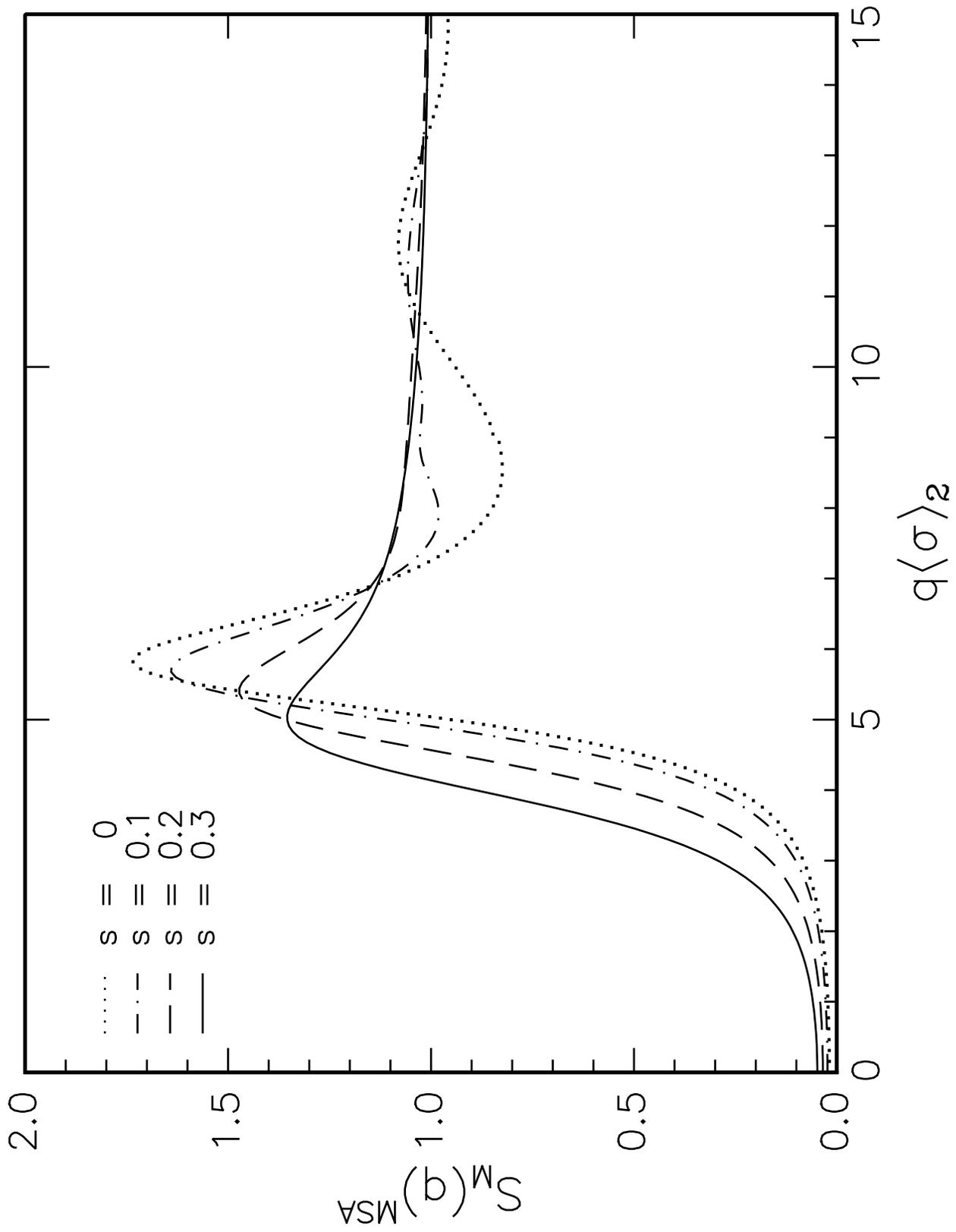
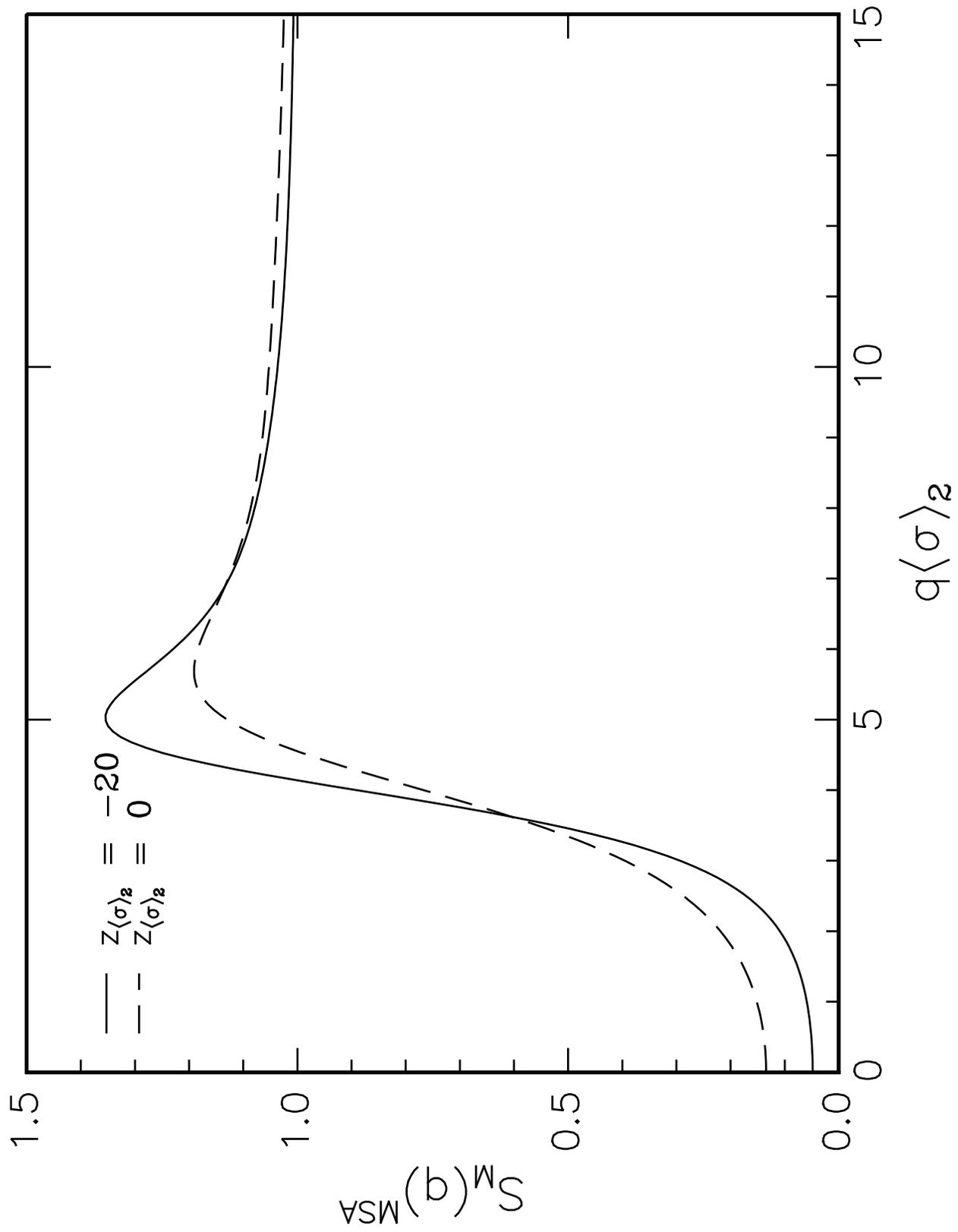
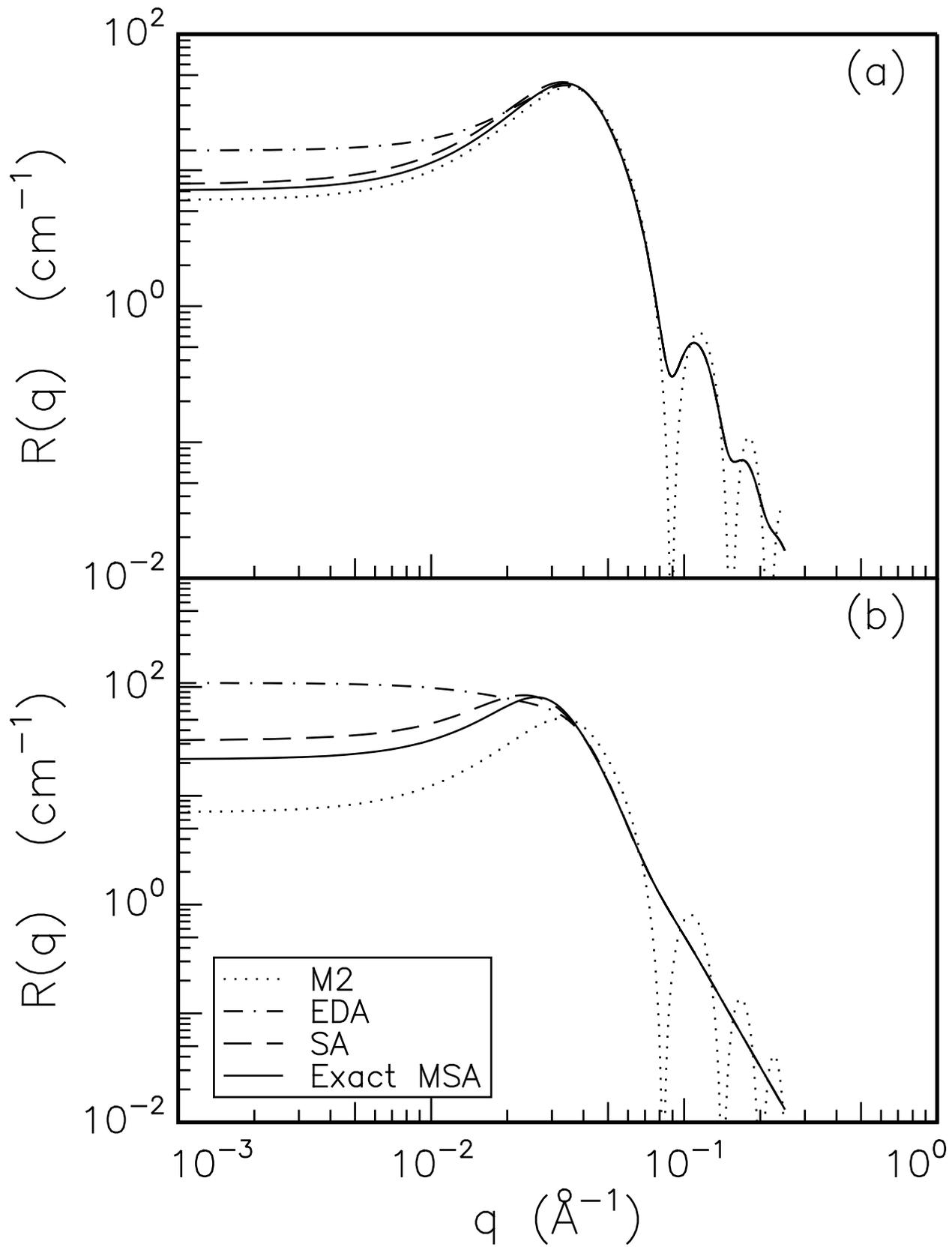
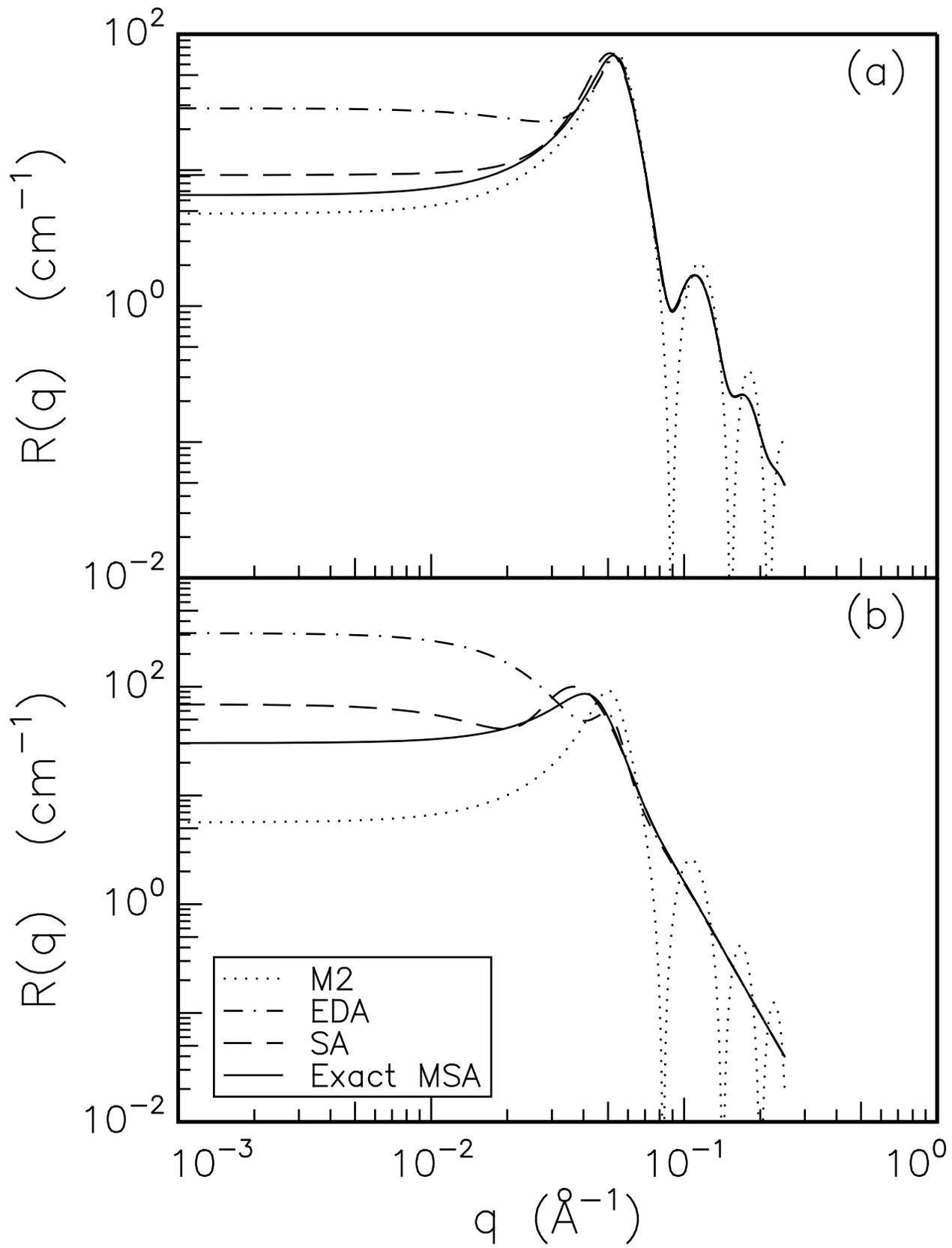
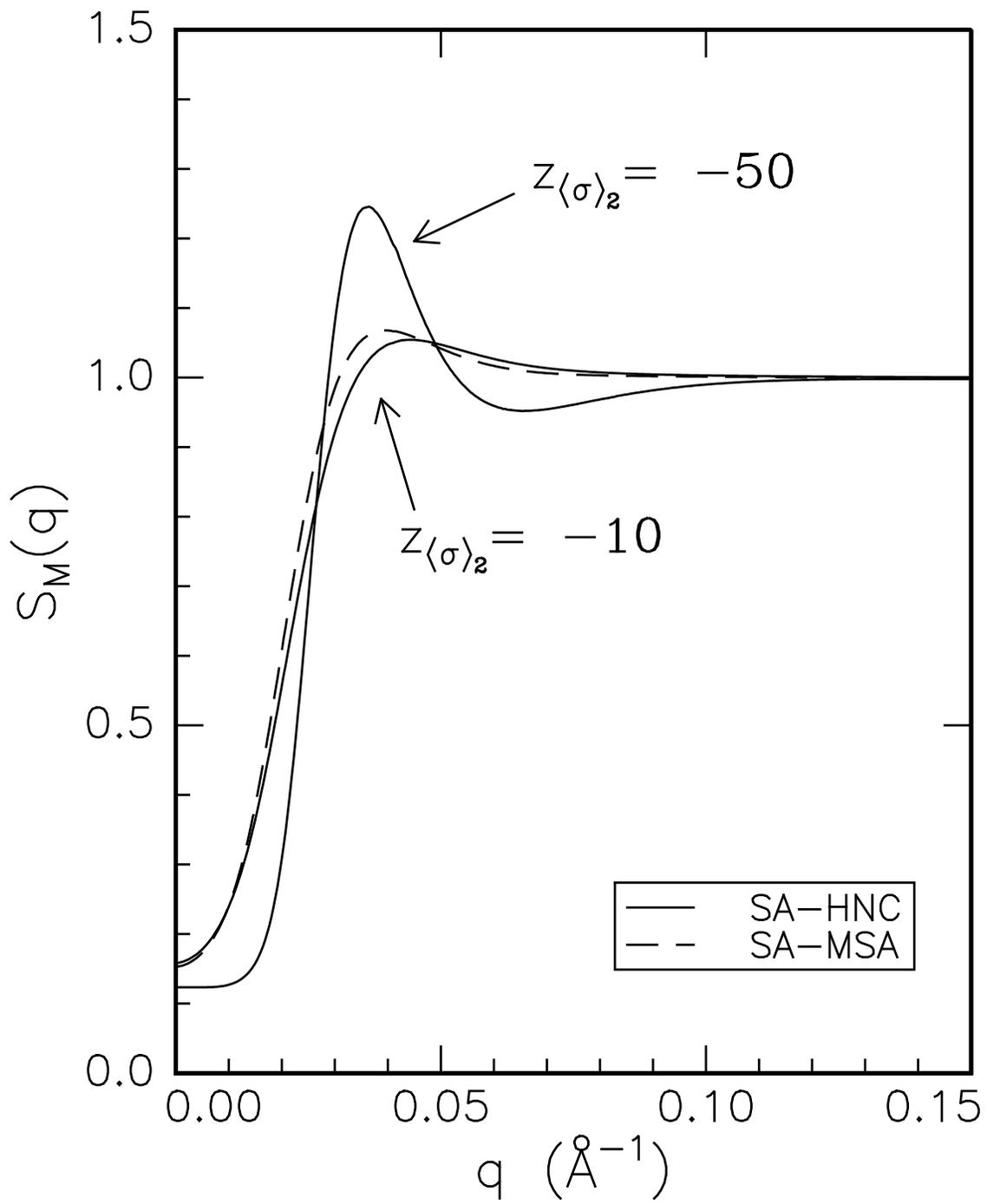
\end{document}